\documentclass[aps,prb,preprint,showpacs]{revtex4}
\usepackage{amssymb}
\usepackage{amsmath}
\usepackage{epsfig}
\usepackage{color}

\newcommand {\cn} {{\rm cn}}

\newcommand {\dn} {{\rm dn}}
\begin{document}

\title{Nonlinear Dynamics in Double Square Well Potential}
\author {Ramaz Khomeriki${}^{1,2}$,  J\'er\^ome Leon${}^3$, Stefano
Ruffo${}^1$, Sandro Wimberger${}^{4}$}
\affiliation {${\ }^{(1)}$Dipartimento di Energetica ``S. Stecco" and CSDC,
Universit\`a di Firenze and INFN, via s. Marta, 3, 50139 Firenze (Italy) \\
${\ }^{(2)}$Department of Exact and Natural Sciences, Tbilisi State University, 3 Chavchavadze, 0128 Tbilisi (Georgia)\\ 
${\ }^{(3)}$ Laboratoire de Physique Th\'eorique et Astroparticules CNRS-UMR5207, Universit\'e Montpellier 2, 34095 Montpellier (France)\\ 
${\ }^{(4)}$CNR-INFM and Dipartimento di Fisica ``E. Fermi'', 
Universit\`a degli Studi di Pisa, Largo Pontecorvo 3, 56127 Pisa, Italy}

\begin{abstract}   Considering the coherent nonlinear dynamics in double square well potential we find the example of {\it
coexistence} of Josephson oscillations with a self-trapping regime.  This
macroscopic bistability is explained by proving analytically  the simultaneous
existence of symmetric, antisymmetric and asymmetric stationary  solutions of
the associated  Gross-Pitaevskii equation.  The effect is illustrated and
confirmed by numerical simulations. This property allows to make suggestions on possible experiments using Bose-Einstein condensates in engineered optical lattices or weakly coupled optical waveguide arrays. 

\end{abstract}

\maketitle

\section{Introduction}

The problem of nonlinear dynamics in double well potential has been first addressed by Jensen \cite{jensen}  who considered light power spatial oscillations in two coupled nonlinear waveguides, which resemble Josephson
oscillations \cite{joseph,anderson} in the {\em spatial domain}. The latter macroscopic quantum tunneling effect, originally discovered in superconducting junctions, is caused by the global phase coherence between electrons in the different layers. More recently the similar realization of a bosonic Josephson junction has been reported for a Bose-Einstein condensate embedded in a macroscopic double harmonic well potential \cite{ober}. The difference with the ordinary Josephson junction behavior is that the oscillations of atomic population imbalance are suppressed for high imbalance values and a self-trapping regime emerges \cite{smerzi1,smerzi2}.

The nonlinear dynamics of bosonic junctions, described by the Gross-Pitaevskii
equation (GPE) \cite{gros}, is usually mapped to a simpler system characterized
by two degrees of freedom (population imbalance and phase difference) while the
nonlinear properties of  the wave function within the single well are
neglected. In this approach the symmetric and antisymmetric stationary
solutions of GPE are used as a basis to build a global wave function
\cite{kivshar1,ananikian}. This description allows to show that for higher
nonlinearities the symmetric solutions become unstable and degenerate to an
asymmetric stationary (approximate) solution of the GPE corresponding to a new
self-trapping regime \cite{alberto,kevrekidis}.

On the other hand, considering the double square well potential (instead of harmonic one), we discover that, in a wide
range of nonlinearities, the system can either remain trapped mostly in one of the wells, or swing periodically from right to left and back. The switching from one state to the other is triggered by a slight local variation of the potential barrier between the wells. The {\it coexistence} of oscillatory and self-trapping regimes corresponds to the simultaneous presence of  Josephson oscillations and of an asymmetric solution of the GPE. 

Our result differs from known behaviors of bosonic Josephson junctions, where
the presence of oscillatory or self-trapping regimes is uniquely determined by
the parameters of the system. The resulting switching property 
should have a straightforward experimental realization in waveguide arrays, which constitute truly one-dimensional systems and are particularly convenient for the observation of nonlinear effects \cite{eisenberg,morandotti,mandelik,yuri,fleisher1,
fleisher2,assanto1} and in engineered optical lattices of Bose-Einstein condensates \cite{ober,smerzi1,smerzi2,ananikian,alberto,cata,min}.

\section{Exact Nonlinear Solutions in Double Square Well}

\begin{figure}[t] \epsfig {file=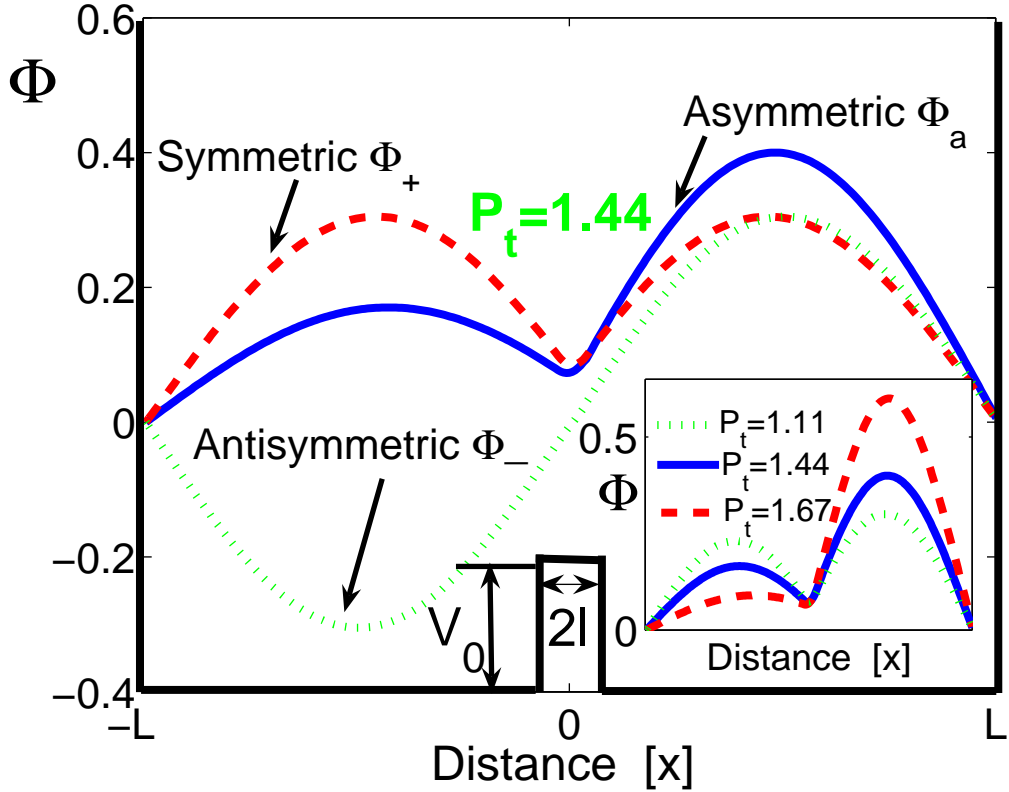,width=0.9\linewidth} 
\caption {Plot of the double square well potential for the
continuous model \eqref{3}: $2L$ is the well width, $V_0$ and $2l$ are barrier
height and width. The curves are the plots of different types of solutions
obtained for the total power $P_t=1.44$.  The inset shows the form of the
asymmetric solution for different values of the total power.}\label{fig:0.25}
\end{figure}

Let us write GPE with double square well potential as follows:
\begin{equation}
i\frac{\partial \psi}{\partial z}+\frac{\partial^2 \psi}{\partial x^2}-V(x)
\psi+|\psi|^2\psi=0, \label{3}
\end{equation}   
where  $V(x)$ is the double square well (represented in Fig.~\ref{fig:0.25}) with a total width $2L$ and the potential barrier height and width $V_0$ and
$2l$, respectively.

The stationary solution of~\eqref{3} are sought as $\psi(z,x)=\Phi(x)
\exp(-i\beta z)$ with a real-valued function $\Phi(x)$ found in terms of
Jacobi elliptic functions \cite{book}
\begin{align}
-L<x<-l\ :\ &\Phi=B\ \cn [\gamma_B(x+L)-{\mathbb K}(k_B),k_B],  \nonumber\\
l<x<L\ :\ &\Phi=A\ \cn [\gamma_A(x-L)+{\mathbb K}(k_A),k_A] , \label{4}\\
-l<x<l\ :\ &\Phi=C\ \dn [\gamma_C(x-x_0),k_C], \nonumber
\end{align}
with the parameters given in terms of the amplitudes by
\begin{eqnarray}
\gamma_A=\sqrt{A^2+\beta}, \quad \gamma_B&=&\sqrt{B^2+\beta}, \qquad
k_A^2=\frac{A^2}{2(A^2+\beta)}, \quad k_B^2=\frac{B^2}{2(B^2+\beta)}, \nonumber \\
\gamma_C^2&=&w-\beta-\frac{C^2}{2}, \quad k_C^2=\frac{V_0-\beta-C^2}{V_0-\beta-C^2/2}, \nonumber
\end{eqnarray}
where ${\mathbb K}$  denotes the complete elliptic integral of the first kind and by construction the above expressions verify the vanishing boundary values in $x=\pm L$.

The solutions are then given in terms of five parameters ($A$, $B$, $C$,
$x_0$, $\beta$), four of which are determined by the continuity conditions in
$x=\pm l$. Thus the conserved total injected power (nonlinearity parameter) $P_t=\int|\psi|^2dx$ completely determines the solutions. Another useful conserved quantity is the total energy $E$ given by
\begin{equation}
E=\int\left(\left|\frac{\partial\psi}{\partial x}\right|^2+
V(x)|\psi|^2-\frac{|\psi|^4}{2}\right)dx. \label{energy}
\end{equation}
In the weakly nonlinear limit (small $P_t$), the solutions are symmetric 
(odd or even). The even solution $\Phi_+(x)$ corresponds to $A=B$ in~\eqref{4}, while odd
solution $\Phi_-(x)$ corresponds to $A=-B$. For higher powers, namely above a threshold value, an
asymmetric solution $\Phi_a(x)$ also exists for which $A\neq \pm B$. These
analytical solutions are represented in Fig.~\ref{fig:0.25}.

\begin{figure}[t]
\epsfig {file=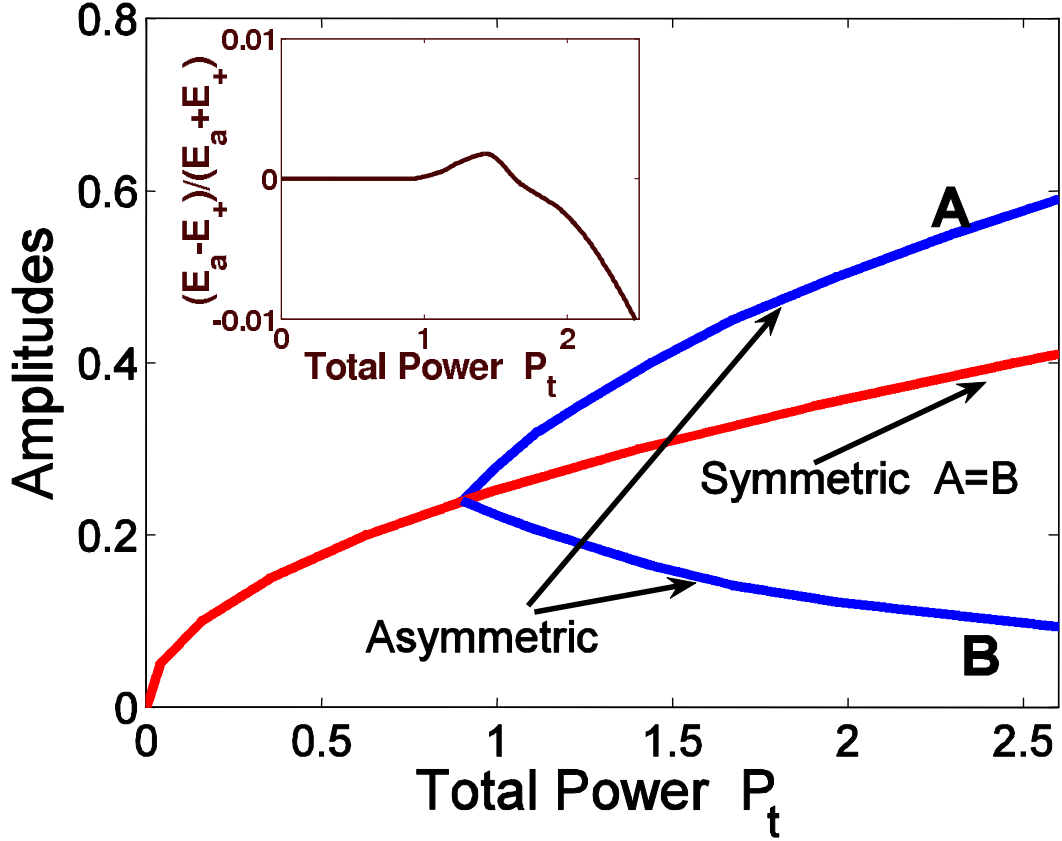,width=0.85\linewidth}
\caption {Dependence of the amplitudes (maximum values $A$ and
$B$ of the expressions \eqref{4}) of the symmetric and asymmetric solutions on
the total power (the amplitudes of the odd and even symmetric solutions almost
superpose). The inset displays the relative energy difference of the symmetric
($\Phi_+$) and assymetric ($\Phi_a$) solutions in terms of the total power.}
\label{fig:bist} \end{figure}

To plot the solutions we stick with the following parameter values: the width of the rectangular double well potential is $2L=7.5$,
the barrier width is $2l=0.25$  and its height is $V_0=20$. We derive the
complete set of solutions \eqref{4} and display the dependence of their
amplitudes on the total power $P_t=\int|\psi|^2dx$ in the main plot of
Fig.~\ref{fig:bist}. Below the threshold value $P_t \approx 0.9$ only the symmetric (odd
and even) solutions exist and their amplitudes almost superpose. At the
threshold value a new solution appears which is asymmetric with amplitudes $A$
and $B$ in the two wells, respectively, represented by the upper  and lower  branches in Fig.~\ref{fig:bist}. 

\section{Two Mode Approximation}

The regime of Josephson oscillation is usually understood on the basis of {\it coupled mode} approach as follows. Using the symmetric and antisymmetric solutions, one builds a variational anzatz by seeking the solution $\psi(z,x)$ under the form
\begin{align}
&\psi(z,x)=\psi_1(z)\Phi_1(x)+\psi_2(z)\Phi_2(x), \label{555} \\
&\sqrt{2}\Phi_1=\Phi_+ +\Phi_-, \quad \sqrt{2}\Phi_2=
\Phi_+ -\Phi_-. \nonumber
\end{align}
The functions $|\psi_1(z)|^2$ and $|\psi_2(z)|^2$ are interpreted as the  
{\em probabilities} to find the system localized either on the left or on the right part of the double square well. 
By construction, the overlap of $\Phi_1$ with $\Phi_2$ is negligible,
consequently, the projection of  the GPE \eqref{3}
successively  on  $\Phi_1$ and $\Phi_2$ provides the coupled mode equations 
\cite{jensen,smerzi1}
\begin{eqnarray}
i\frac {\partial \psi_1}{\partial z}+ D|\psi_1|^2\psi_1=r\psi_{2},
\nonumber \\
i\frac {\partial \psi_2}{\partial z}+ D|\psi_2|^2\psi_2=r\psi_{1},
\label{222} 
\end{eqnarray}
with coupling constant $r$ and nonlinearity parameter $D$ defined by
\begin{equation*}
r=\frac {\int\left[(\partial_x\Phi_1)(\partial_x\Phi_2)+V\Phi_1\Phi_2\right]dx}
{\int\Phi_1^2\,dx}, \quad
D=\frac {\int\Phi_1^4\,dx}{\int\Phi_1^2\,dx}. 
\end{equation*}

An explicit solution of \eqref{222} in terms of Jacobi elliptic functions  has
been found in \cite{jensen} and used in Bose-Einstein condensates in
\cite{smerzi2}. It is a good approximation for the system in a double harmonic
potential well \cite{kevrekidis} and correctly describes the oscillatory regime
in our case. Indeed, when the power is initially injected into one array, say
$|\psi_1(0)|=1$, $|\psi_2(0)|=0$, we obtain for $D<4r$
\begin{equation}
|\psi_1|^2=\frac12\left[1+\cn (2rz, \frac D{4r})\right],\quad 
|\psi_2|^2=1-|\psi_1|^2.
\label{7a}\end{equation}
Since $|\psi_1|$ oscillates  around the value $0$, this expression describes an
oscillation of light intensity between the left and the right wells. The period
of this oscillation is
\begin{equation}
T=2{\mathbb K}(D/4r)/r \label{888}
\end{equation}
and has been checked on various numerical shots at different total input
power. In summary, while the self-trapping regime is directly interpreted in
terms of the asymmetric solution, the interpretation of the Josephson
oscillation regime  needs to call to the coupled mode approach, which in turn
fails to explain the observed {\em coexistence} of both regimes.
\begin{figure}[ht]
\epsfig{file=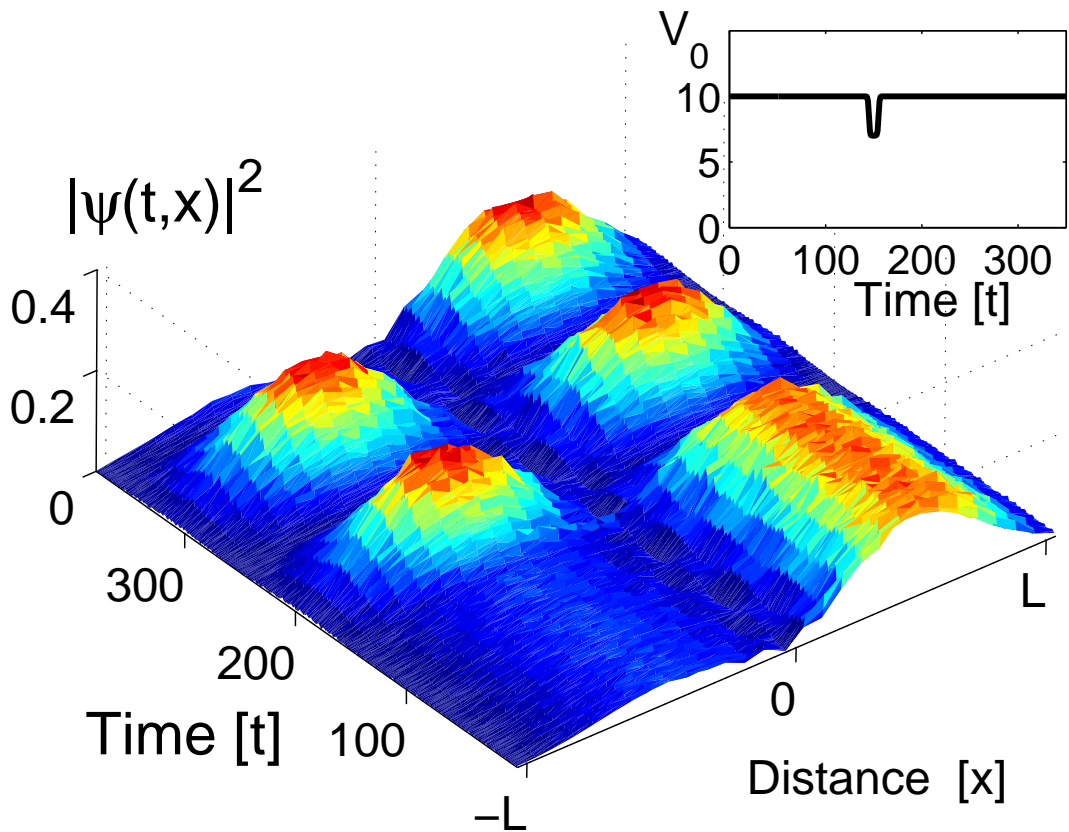,width=\linewidth}  
\caption {Numerical simulation of the GPE equation~\eqref{3}.  By a slight
local variation at $z=150$ of the potential barrier height,
represented in the inset, the regime switches from self-trapping to Josephson oscillations. The injected
total power is $P_t=\sum|\psi_j|^2=1.44$.} 
\label{fig:oscpin}\end{figure}
 
Such a coexistence, however, is understood in terms of the energy \eqref{energy}
which can be evaluated, at given total power $P_t$, both for the 
symmetric solution $\Phi_+$ and for the asymmetric solution
$\Phi_a$. As shown in the inset of Fig.~\ref{fig:bist} these two energy values
$E_+$ and $E_a$ turn out to be very close up to the total power value $P_t\approx
2$. Consequently, switching from a regime to the other is allowed at fixed
power. In particular, in the numerical experiments of Fig.~\ref{fig:oscpin},
total power and energy are the same before and after the local variation of
the potential barrier value.

It is worth to remark that a similar analysis in the case of harmonic double well
potentials \cite{alberto,kevrekidis} shows that the energy of the asymmetric
solution (when this solution exists) is significantly smaller than the energy of the symmetric solution. In such a situation, it is thus impossible to switch from a self-trapped state to an oscillatory regime when keeping both the energy and the total power constant.

\section{Applications for BEC and Coupled Waveguide Arrays}

\begin{figure}[h]
\begin{center} 
\epsfig{file=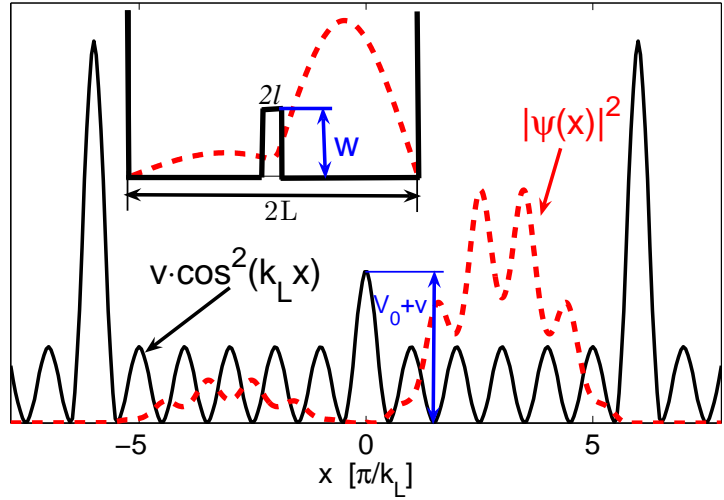,width=0.6\linewidth} 
\end{center} 
\caption{Schematics of the suggested experimental setup. In the context of BEC the optical lattice is supplemented by large energy barriers from 
both sides and a small one in the middle (solid curve). The condensate is initially loaded mainly into the right part of the optical lattice 
(the dashed line represents particle density). The inset shows the reduction of the problem 
to the particle motion in a double square well potential (details are given in the text). In the context of waveguide arrays the solid curve displays (with the opposite sign) variation of the refractive index across the array.} 
\label{becexp}\end{figure}

Now our aim in this section is to suggest the realistic experiments on BEC and waveguide arrays, which are engineered in such a way to
mimic two weakly coupled chains of JJ's (see Fig. \ref{becexp}). Using such an experimental set-up, we demonstrate 
the feasibility of the efficient control of a switch between oscillating and 
self- trapping states of the systems. We show that our problem reduces to the Gross-Pitaevskii equation 
(GPE) \cite{gros} in a double square well, which displays very different properties from the previously considered double
harmonic well potential \cite{ober,smerzi1,smerzi2,alberto,kevrekidis,min}.
Our results are broadly applicable and open the way to
the experimental study of these phenomena in the dynamics of those systems.

We start the consideration of the case of BEC in an optical lattice, for which a  one-dimensional Hamiltonian has a following form: 
\begin{equation}
i\hbar\frac{\partial \psi}{\partial t}=-\frac{\hbar^2}{2m}\frac{\partial^2 \psi}{\partial x^2}+V(x)\psi+\frac{2\hbar^2a_s}
{ma_\perp^2}|\psi|^2\psi, \label{122}
\end{equation} 
where $m$ is atomic mass, $a_s<0$ is the scattering length corresponding to the attractive atom-atom interactions and 
$a_\perp=\sqrt{\hbar/m\omega_\perp}$ is the transversal oscillation length,  which implicitly takes
into account the real three dimensionality of the system \cite{equation}, $\omega_\perp$ being the transversal frequency of the trap. 
The optical lattice potential is
\begin{eqnarray}
V(x)=v\cos^2(k_Lx) \qquad \mbox{for} \qquad |k_Lx|>\pi/2 \nonumber \\ V(x)=\bigl(v+V_0\bigr)\cos^2(k_Lx) \qquad \mbox{for} \qquad |k_Lx|<\pi/2, 
\label{111}
\end{eqnarray}
where $k_L$ is the wavenumber of the laser beams that create the optical lattice and $V_0$ is the height of the additional 
spatial energy barrier placed in the middle of the optical lattice. Besides that, Dirichlet boundary conditions with 
$\psi(\pm L)=0$ are chosen in order to describe the large confining barriers at both ends of the BEC. These boundary 
conditions could be realized experimentally by an additional optical lattice with larger amplitude and larger lattice constant, as shown in Fig.\ref{becexp}.   

Introducing a dimensionless length scale $\tilde x=2k_Lx$ and time $\tilde t=E_B t/\hbar$, where $E_B=8E_R=4\hbar^2k_L^2/m$ and $E_R$ is the 
recoil energy \cite{bloch}, we can rewrite \eqref{1} as follows
\begin{equation}
i\frac{\partial \Psi}{\partial \tilde t}=-\frac{1}{2}\frac{\partial^2 \Psi}{\partial \tilde x^2}+\tilde V(\tilde x)\Psi+g|\Psi|^2\Psi, \label{2}
\end{equation}
where the normalized wave-function, $\int |\Psi(\tilde x)|^2d\tilde x=1$, is  introduced \cite{Schlagheck}. 
The dimensionless potential $\tilde V$ still has the form \eqref{111} with the following dimensionless depths of the optical lattice
\begin{equation}
\tilde v=\frac{v}{E_B}, \qquad \tilde V_0=\frac{V_0}{E_B}, \qquad
g=\frac{Na_s}{k_La_\perp^2}\;, 
\label{333}
\end{equation}
$g$ being the dimensionless nonlinearity parameter.

We have performed numerical simulations of Eq. \eqref{2} with 12 wells (6 wells on each side of the barrier as presented in Fig. \ref{becexp}) 
and the parameters $\tilde v=0.25$ (in physical units this means that the depth of the optical lattice is $v=2E_R$), $\tilde V_0=0.15$ 
and we fix the nonlinearity to the value $g=-0.025$, i.e., we choose attractive interactions. The dynamics is similar for repulsive interatomic 
forces (see the discussion below). The phenomenon we study in this Letter does not depend significantly on the actual size of the system, if at least 3 lattice sites are present at each side of the barrier.
\begin{figure}[ht]
\begin{center}
\epsfig{file=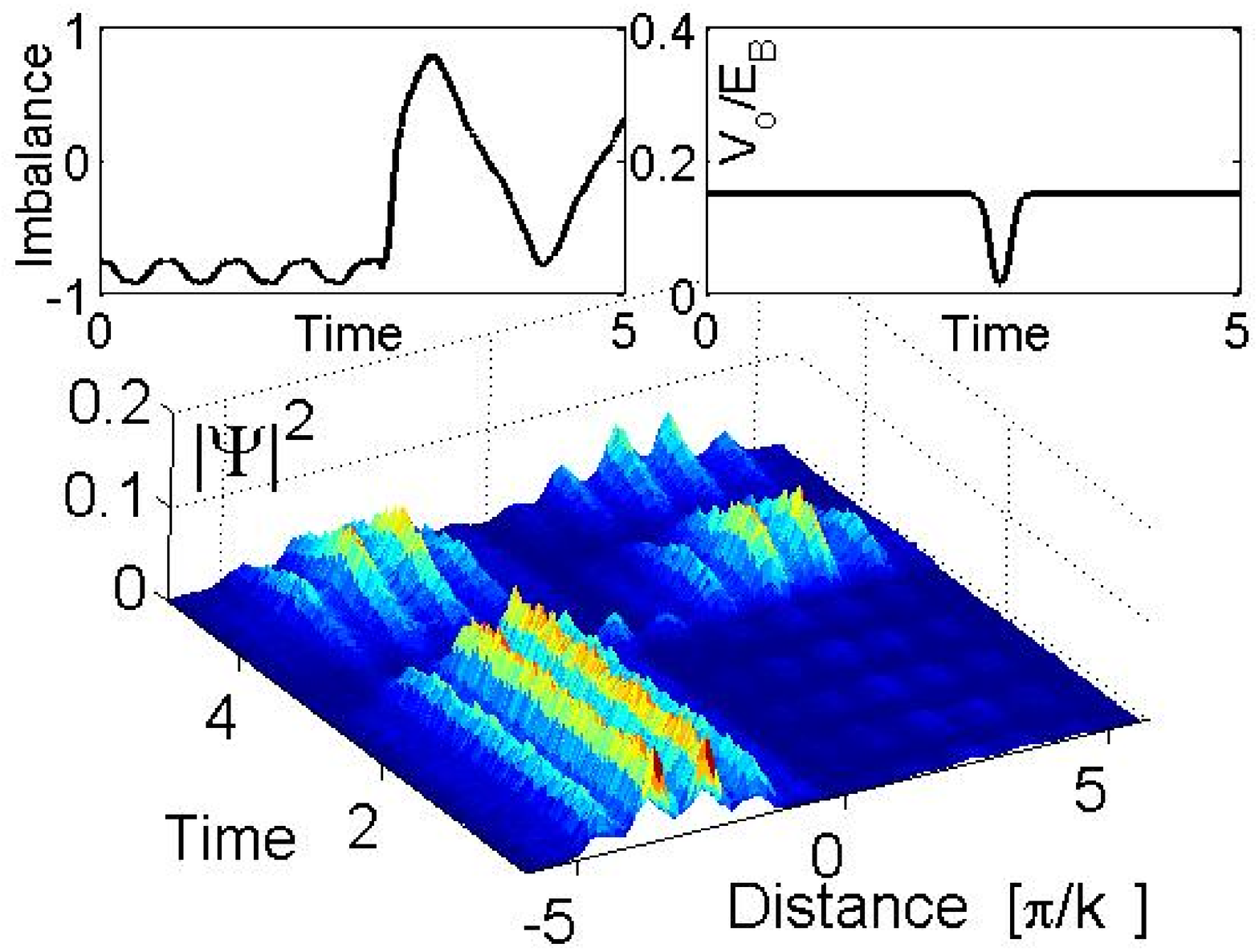,angle=90,width=0.49\linewidth} 
\epsfig{file=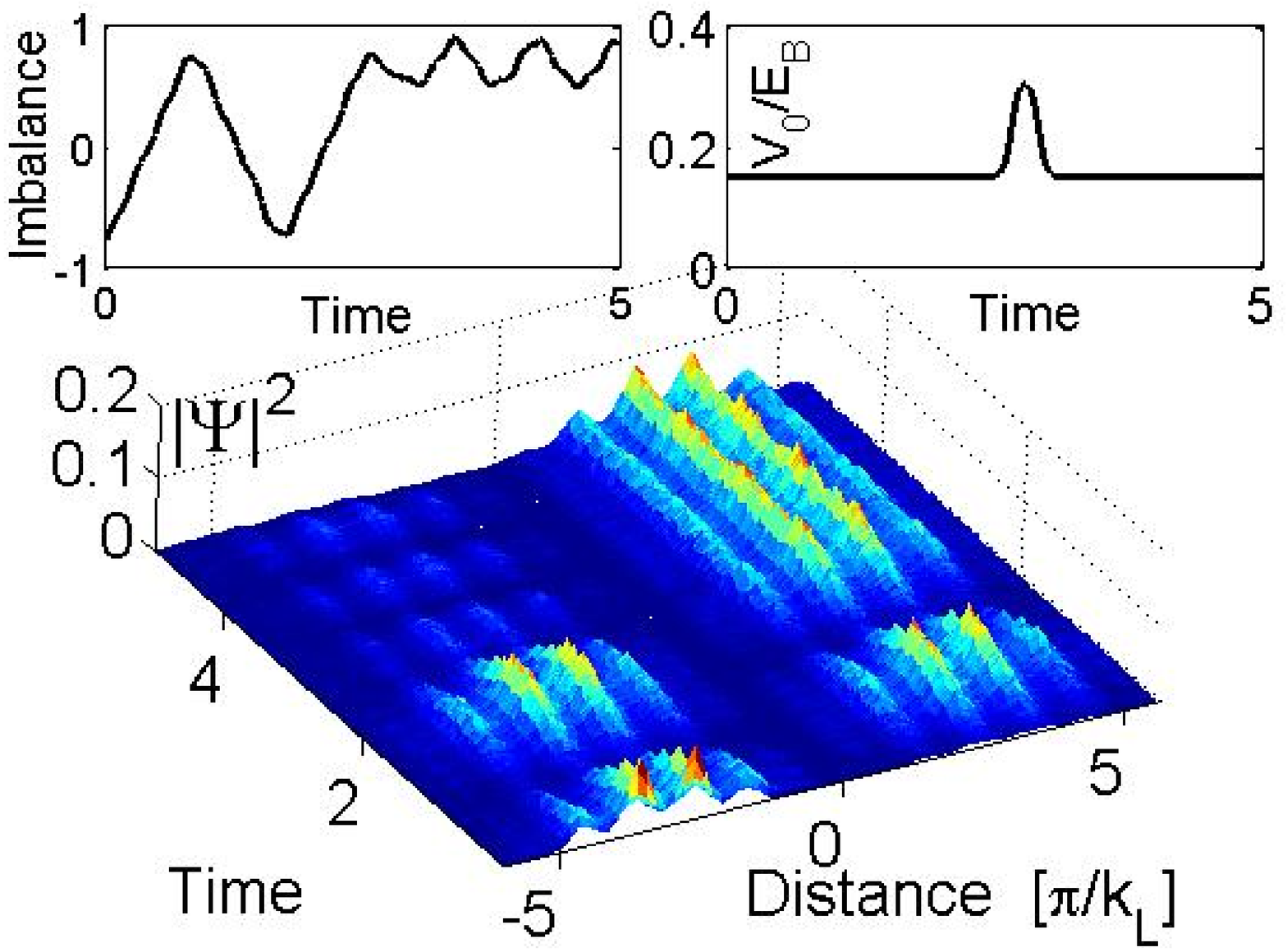,angle=90,width=0.49\linewidth}
\end{center} 
\caption{Numerical simulations of Eq. \eqref{2}: the left graph represents the transition from a self-trapped 
state to the macroscopic tunneling regime, while the right graph describes the inverse process. The insets in both graphs 
show the variation of the energy barrier necessary to realize the switching between the different regimes.} 
\label{becosc}\end{figure}

As seen from the left panel of Fig. \ref{becosc}, if one prepares the condensate in a self-trapped state it remains there until we 
apply the pulselike time variation of the barrier displayed in the inset. After that action, the condensate goes into the oscillating 
tunneling regime. On the other hand, preparing the condensate in the oscillating tunneling regime (right graph in Fig. \ref{becosc}) 
one can easily arrive at a self-trapping state by varying again the energy barrier in the middle as displayed in the inset. Let 
us mention that, as far as the energy of the barrier is changed adiabatically, the total energy of the condensate does not vary, 
i.e. the self-trapped and tunneling oscillatory regimes have the {\em same} energy. This is quite different from what happens 
in a double harmonic well potential \cite{ober,smerzi1,smerzi2,alberto,kevrekidis,min}. The point is that, in the double harmonic well, 
the asymmetric stationary solution is characterized by a smaller energy than the symmetric solution and this difference 
increases sharply with increasing nonlinearity. Hence, a drastic energy injection is required in order to realize the transition between the 
two regimes; whilst in our case the transition is simply achieved only by varying pulsewise the energy barrier. 
Below we argue that this happens because our case effectively reduces to the case of a double square well potential (see the inset 
of Fig. \ref{becexp} and the reduction procedure below) for which asymmetric and symmetric stationary solutions carry almost 
the same energies in a wide range of the nonlinearity parameter.  

Now we proceed to reducing Eq. \eqref{2} to a Discrete NonLinear Schr\"odinger equation (DNLS). We discretize it via 
a tight-binding approximation \cite{smerzitight,yuri,mark}, representing the wave function $\Psi(\tilde x)$ as
\begin{equation}
\Psi(\tilde x)=\sum_j\phi_j\Phi_j(\tilde x), 
\label{6}
\end{equation}
where $\Phi_j(\tilde x)$ is a normalized isolated wave function in an optical lattice in the fully linear case $g=0$ and could 
be expressed in terms of Wannier functions (see, e.g., \cite{wanier}). For clarity, we use here its approximation for a harmonic 
trap centered at the points $r_j=j\pi(|j|+1/2)/|j|$ ($|j|$ varies from 1 to $n$, the number of wells). In the context of the 
evolution equation \eqref{2} $\Phi_j(\tilde x)$ has the form
\begin{equation}
\Phi_j(\tilde x)=\left(\frac{\sqrt{\tilde v}}{\pi\sqrt{2}}\right)^{1/4}e^{-\sqrt{\tilde v}(\tilde x -r_j)^2/\sqrt{8}}  \;,
\label{7}
\end{equation}
for $|j|\neq 1$, and one should substitute $\tilde v$ by $\tilde v+\tilde V_0$ in the above expression in order to get an 
approximate formula for the wave function for $|j|=1$.

Assuming further that the overlap of the wave functions in neighboring sites is small, we get from \eqref{2} the following DNLS equation for the sites $|j|\neq 1$
\begin{equation}
i\hbar\frac{\partial \phi_j}{\partial \tilde t}=-Q\bigl(\phi_{j+1}+\phi_{j-1}\bigr)+U|\phi_j|^2\phi_j, \label{8}
\end{equation}
while for $|j|=1$ we have
\begin{equation}
i\hbar\frac{\partial \phi_{\pm 1}}{\partial \tilde t}=-Q\phi_{\pm 2}-Q_1\phi_{\mp 1}+U_1|\phi_{\pm 1}|^2\phi_{\pm 1}, \label{81}
\end{equation}
where we assume pinned boundary conditions. The constants $Q$, $Q_1$, $U$ and $U_1$ are easily computed from the following expressions ($|j|\neq 0$):
\begin{eqnarray}
Q&=&-\int\left[\frac{\partial\Phi_j}{\partial \tilde x}\frac{\partial\Phi_{j+1}}{\partial \tilde x}+\tilde v\cos^2(\tilde x/2)\Phi_j\Phi_{j+1}\right]d\tilde x, \nonumber \\
Q_1&=&-\int\left[\frac{\partial\Phi_1}{\partial \tilde x}\frac{\partial\Phi_{-1}}{\partial \tilde x}+(\tilde v+\tilde V_0)\cos^2(\tilde x/2)\Phi_1\Phi_{-1}\right]d\tilde x, \nonumber \\
U&=&g\int\Phi_j^4\,d\tilde x\simeq U_1=g\int\Phi_{\pm 1}^4\,d\tilde x \;.
\label{9} 
\end{eqnarray}
In order to characterize the solutions of Eqs. \eqref{8} and \eqref{81}, we follow the same procedure used in Ref. \cite{jerome}, which
goes through a continuum approximation. Assuming that $\phi_1=\phi_{-1}$ we finally arrive at
\begin{equation}
\frac{i\hbar}{Q}\frac{\partial \phi(j)}{\partial \tilde t}=-\frac{\partial^2 \phi(j)}{\partial j^2}+W(j)\phi(j)+ R|\phi(j)|^2\phi(j), \label{10}
\end{equation}
where now $j$ is a continuous variable, $W(j)$ is a double square well potential with a barrier height $w=2(Q-Q_1)/Q$ and width $l=1$, 
$\phi(j)$ obeys pinned boundary conditions $\phi(j=\pm L)=0$ ($2L$ is a width of a double square well potential) and the nonlinearity parameter 
is given by $R= U/Q<0$. Expressing $\psi\equiv \sqrt{|R|}\phi$ and $z\equiv Q\tilde t/\hbar$ and mentioning that total power $P_t$ is connected with nonlinearity parameter $R$ as $P_t=|R|$, we see the equation \eqref{10} is the same as \eqref{3} and thus all the above consideration of peculiarities of double square well potential directly applies to the considered BEC lattices.

In case of the waveguide systems the situation is even simpler. Particularly, as well known an array of adjacent waveguides coupled by power exchange is modeled by the discrete nonlinear Schr\"odinger equation (DNLS)~\cite{christ-joseph,mark}
which reads
\begin{equation}
i\frac{\partial \psi_j}{\partial z}+\frac{\omega}
{c}(n_j-n)\psi_j+Q\bigl(\psi_{j+1}+\psi_{j-1}-2\psi_{j}\bigr)+ 
|\psi_j|^2\psi_j=0,
 \label{1} \end{equation}
where waveguides discrete positions are labelled by the index $j$  ($-N\leq j
\leq N$), and the complex field $\psi_j$ results from the projection of the
electric field envelope on the eigenmode of the individual waveguide. It is
normalized to a unit onsite nonlinearity. The linear refractive index $n_j$ is
set to $n$ for all  $j\ne 0$, and to $n_0<n$ for $j=0$. The coupling constant
between two adjacent waveguides is $Q$ and $\omega$ and $c$ are the light
frequency and velocity. Vanishing boundary conditions
$\psi_{N+1}=\psi_{-N-1}=0$ model a strongly evanescent field outside the
waveguides. Considering now $1/\sqrt Q$ as being a virtual grid spacing we
may represent $\psi_j(z)$ by the function $\psi(x,z)$ in the continuous variable
$x=j/\sqrt{Q}$. As a result the DNLS model \eqref{1} maps to the \eqref{3} with a double square well potential considered initially.

\section{conclusions}

A new coherent state in square double well potential has been discovered.
This coherent state has the property of being {\it bistable}: one can easily
switch from oscillatory to self-trapping regimes and back. This nontrivial
behavior may have interesting applications in various weakly linked extended
systems, such as Bose-Einstein condensates, waveguide or Josephson junctions arrays, which deserve further studies.

In the region of nonlinearities where the asymmetric solution coexists with 
the symmetric and asymmetric stationary solutions, we have induced the switch
from one regime to the other by varying the height of the barrier. In view of a real experiment 
one could induce such flips by varying the refractive index of the central waveguide (in the context of weakly linked waveguide arrays) or by pulswize change of optical barrier potential (in case of BEC).

{\bf Acknowledgements:} We would like to thank F.T. Arecchi, E. Arimondo, A. Montina and O. Morsch for useful discussions. R. Kh. acknowledges support by
Marie-Curie international incoming fellowship award (MIF1-CT-2005-021328) and
NATO grant (FEL.RIG.980767), S. R. acknowledges financial support under the
PRIN05 grant on {\it Dynamics and thermodynamics of systems with long-range
interactions} and S.W. is funded by the Alexander von Humboldt foundation (Feodor-Lynen Program).

\end{document}